\documentclass[pra,twocolumn,notitlepage,showpacs]{revtex4-1}

\usepackage{bbm}
\usepackage{amsmath}
\usepackage{verbatim}
\usepackage{graphicx}
\usepackage{times}
\usepackage{amssymb}
\usepackage{epsfig}
\usepackage{graphicx}
\usepackage{bm}
\usepackage{times}
\usepackage{txfonts}
\usepackage{dsfont}
\usepackage{amsfonts}

\begin{document}

\title{Optical quantum swapping in a coherent atomic medium}

\author{Aur\'elien Dantan}

\affiliation{QUANTOP, Danish National Research Foundation Center for Quantum
Optics,\\ Department of Physics and Astronomy, University of Aarhus,
DK-8000 Aarhus C, Denmark
}

\begin{abstract}
We propose to realize a passive optical quantum swapping device which allows for the exchange of the quantum fluctuations of
two bright optical fields interacting with a coherent atomic medium in an optical cavity. The
device is based on a quantum interference process between the fields within the cavity
bandwidth arising from coherent population trapping in the
atomic medium.
\end{abstract}


\maketitle

Manipulating the quantum state of bright optical fields is at the core of quantum information
processing in the continuous variable regime~\cite{braunstein05,QICV,ralph09}. However, even
basic linear processing tasks, such as the quantum state swapping between two different frequency fields, can be challenging to achieve in practice. Unconditional exchange of quantum fluctuations
can be achieved for instance using complex teleportation and entanglement swapping protocols~\cite{braunstein98,furusawa98,polkinghorne99,takei05}, which typically require entanglement and/or measurements combined with active feedback~\cite{hammerer10,furusawa11}. We propose here to realize a passive quantum swapping device which allows for an efficient exchange of the quantum fluctuations of two optical beams interacting in an optical cavity with a coherent atomic medium consisting of three-level $\Lambda$ atoms. In a coherent population trapping (CPT) situation, when the two fields are resonant with the ground-to-excited state transitions and strongly drive the atoms into a coherent superposition
of the two ground-states, the medium becomes transparent for the fields~\cite{arimondo76}. Like in Electromagnetically Induced Transparency (EIT) this behavior occurs within a certain frequency window around two-photon resonance defined by the effective cavity linewidth $\kappa_{CPT}$, which, for a sufficiently high effective optical depth of the medium, can be much narrower than that of the bare cavity $\kappa$~\cite{lukin98,hernandez07,wu08,mucke10,kampschulte10,albert11}. We show that, like the field classical mean-values, the quantum fluctuations are also preserved within this transparency window. However, in the frequency range $\kappa_{CPT}<\omega<\kappa$, where $\omega$ is the frequency of the sidebands considered, both fields are shown to exchange their respective fluctuations, thus realizing a quantum swapping operation. This exchange arises from the CPT-induced quantum interferences which affect in a different way the dark and bright field mode combinations which are uncoupled and coupled, respectively, with the atomic medium. We show that, for a wide range of parameters, the system can act as a lossless, frequency-dependent phase-plate for the field sidebands and achieve efficient quantum state swapping.

Quantum interference effects due to coherent population trapping or electromagnetically induced transparency, in particular the strong dispersion and low absorption that can be experienced by the fields, have been exploited in various contexts, e.g. for atomic clock spectroscopy~\cite{vanier05}, magnetometry~\cite{budker02}, nonlinear and
quantum optics~\cite{fleischhauer05}. In connection with the present work, it was predicted in~\cite{harris93,agarwal93,fleischhauer94} that the free-space propagation in a resonant CPT medium would generate pulses with matched statistics. Subsequently, the propagation of nonclassical quantum fluctuations in a coherent atomic medium under EIT or CPT conditions was investigated, both theoretically~\cite{fleischhauer02,dantan05,barberis07,hu11} and experimentally~\cite{akamatsu04,hsu06,appel08,honda08,cviklinski08,arikawa10}. The quantum properties of light fields interacting with a coherent medium placed in an optical cavity was also analyzed in connection with quantum memory~\cite{lukin00,dantan04,dantan04epl} and entanglement and spin-squeezing generation~\cite{dantan06}. Here, the role of the cavity with respect to quantum state swapping is double, as it enhances the effective optical depth of the medium - and thereby the swapping efficiency - as well as provides a passive mechanism for the exchange of the fluctuations in a well-defined frequency range within the cavity bandwidth. Since it is based on a quantum interference effect intrinsically occurring between the fields in the atomic medium the device does not require additional quantum resources, such as entanglement, measurement or active feedback. The proposed mechanism could also have applications in the microwave domain, e.g. with superconducting artificial atoms~\cite{kelly10,bianchetti10}.

\begin{figure}[h]
\center{\includegraphics[width=0.8\columnwidth]{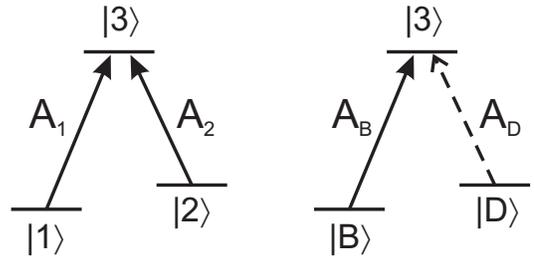}}
\caption{Atomic level structure considered: (a) Initial basis. (b) Dark/bright state basis.}
\label{fig:level}
\end{figure}

We consider an ensemble of $N$ three-level $\Lambda$ atoms with long-lived ground/metastable states
$|1\rangle$ and $|2\rangle$ and excited state $|3\rangle$. The atoms are positioned in an optical
cavity, where they interact with two fields, $A_1$ and $A_2$, resonant with the
$|1\rangle\longrightarrow|3\rangle$ and $|2\rangle\longrightarrow|3\rangle$ transitions,
respectively (fig.~\ref{fig:level}). The cavity is assumed single-ended, lossless and resonant with both fields. Denoting by $P_{ij}$ ($i,j=1-3$) the collective atomic operators, the interaction Hamiltonian in the rotating-wave approximation and in the rotating frame reads \begin{equation} H_{int}=-\hbar(g_1A_1P_{31}+g_2A_2P_{32}+h.c.),\end{equation} where the $g_i$'s are the single-atom coupling strength. The evolution of the atom-field system is given by a set of Heisenberg-Langevin equations (see e.g.~\cite{dantan04,dantan06})
\begin{align*}
\dot{P}_{13}&=-\gamma P_{13}+ig_1A_1(P_{11}-P_{33})+ig_2A_2P_{12}+F_{13},\\
\dot{P}_{23}&=-\gamma P_{23}+ig_2A_2(P_{22}-P_{33})+ig_1A_1P_{21}+F_{23},\\
\dot{P}_{12}&=-\gamma_0 P_{12}-ig_1A_1P_{32}+ig_2A_2^{\dagger}P_{13}+F_{12},\\
\dot{P}_{11}&=\gamma_1P_{33}+ig_1A^{\dagger}_1P_{13}-ig_1A_1P_{31}+F_{11},\\
\dot{P}_{22}&=\gamma_2P_{33}+ig_2A^{\dagger}_2P_{23}-ig_2A_2P_{32}+F_{22},\\
\dot{A}_1&=-\kappa_1A_1+ig_1P_{13}+\sqrt{2\kappa_1}A_1^{in},\\
\dot{A}_2&=-\kappa_2A_2+ig_2P_{23}+\sqrt{2\kappa_2}A_2^{in},
\end{align*}
where $\gamma=(\gamma_1+\gamma_2)/2$ is the optical dipole decay rate, $\gamma_0$ the ground-state coherence decay rate ($\gamma_0\ll\gamma_{1,2}$), $\kappa_1$ and $\kappa_2$ the intracavity field decay rates. The $F_{ij}$'s are zero-mean valued Langevin noise operators, whose correlation functions can be calculated from the quantum regression theorem~\cite{courty91,hilico92}. $A_1^{in}$ and $A_2^{in}$ are the input field operators. By means of the standard linearized input-output theory~\cite{courty91,hilico92}, one can calculate from these equations the mean values of the observables in steady-state (denoted by $\langle \xi\rangle$), and derive the evolution equations of the quantum fluctuations $\delta \xi=\xi-\langle \xi\rangle$. Going to the Fourier space, it is then possible to relate the fluctuations of the fields exiting the cavity, $A_{j}^{out}=\sqrt{2\kappa_{j}}A_{j}-A_{j}^{in}$ $(j=1,2)$, to those of the incoming fields. In particular, one can compare their quadrature noise spectra $S_{X_{j,\theta}}(\omega)$ at a given sideband frequency $\omega$, where the quadrature fluctuations and noise spectra are standardly defined by $\delta X_{j,\theta}=\delta A_j e^{-i\theta}+\delta A_j^{\dagger}e^{i\theta}$ and \begin{equation}\langle \delta X_{j,\theta}(\omega)\delta X_{j,\theta}(\omega')\rangle=2\pi \delta(\omega+\omega')S_{X_{j,\theta}}(\omega)\hspace{0.3cm}(j=1,2).\end{equation}

This full quantum mechanical calculation can be performed without approximation for any Gaussian input field states. However, for the sake of the discussion and in order to derive analytical results, we will in the following focus on the symmetric situation of fields with comparable intracavity Rabi frequencies, $\Omega_{i}=g_i\langle A_i\rangle\sim \Omega$ $(i=1,2)$, sufficient to saturate the two-photon transition: $\Omega\gg\gamma\gamma_0$. In this case, the atoms are pumped into a dark state $|-\rangle=(|1\rangle-|2\rangle)/\sqrt{2}$ with maximal coherence $\langle P_{12}\rangle\sim-N/2$. It is then convenient to turn to the dark/bright state basis ${|-\rangle,|+\rangle}$, where $|+\rangle=(|1\rangle+|2\rangle)/\sqrt{2}$, and define \textit{bright} and \textit{dark} optical modes $A_{\pm}=(A_1\pm A_2)/\sqrt{2}$. The dark mode has then zero mean value, $\langle A_-\rangle=0$, and one finds the situation analyzed in~\cite{dantan04}, in which the atoms, all in state $|-\rangle$, are coupled to the empty dark mode $A_-$ on the transition $|-\rangle\longrightarrow|3\rangle$, and the bright mode $A_+$ on the transition $|+\rangle\longrightarrow|3\rangle$. The bright mode "sees" no atoms, but induces electromagnetic transparency for the dark mode. One can then show that the equations for the fluctuations of the dark mode $A_-$, of the ground-state coherence $Q=|-\rangle\langle +|$ and of the dark dipole $P_-=|-\rangle\langle 3|$ are decoupled from those of the bright mode $A_+$, and given by~\cite{dantan04} \begin{eqnarray}
(\kappa-i\omega)\delta A_-&=&ig\delta P_-+\sqrt{2\kappa}\delta A_-^{in},\\
(\gamma-i\omega)\delta P_-&=&i\Omega'\delta Q+igN\delta A_-+F_-,\\
(\gamma_0-i\omega)\delta Q&=&i\Omega'\delta P_-+F_Q,
\end{eqnarray}
where $\Omega'=\Omega\sqrt{2}$, we assumed $g=g_1=g_2$, $\kappa_1=\kappa_2=\kappa$, $F_-$ and $F_Q$ are zero-mean valued Langevin operators with correlation functions $\langle F_-(\omega)F_-^{\dagger}(\omega')\rangle=2\gamma N\delta(\omega+\omega')$ and $\langle F_Q(\omega)F_Q^{\dagger}(\omega')\rangle=2\gamma_0 N\delta(\omega+\omega')$. Assuming a small ground-state decoherence rate ($\gamma_0\ll\gamma,\kappa$), the fluctuations of the outgoing dark mode are readily found to be \begin{equation}\label{eq:A-out} \delta A_{-}^{out}=\frac{\kappa+i\omega-\beta}{\kappa-i\omega+\beta}\delta A_-^{in}+F_{in},\end{equation} where
\begin{eqnarray} \beta(\omega)=\frac{g^2N(\gamma_0-i\omega)}{(\gamma-i\omega)(\gamma_0-i\omega)+\Omega'^2},\hspace{0.2cm}
F_{in}=\frac{\sqrt{2\kappa}}{\kappa-i\omega+\beta}F_-.\label{eq:F-}
\end{eqnarray} Since the transmission function of the bright mode fluctuations is that of an empty cavity, \begin{equation}\label{eq:A+out} \delta A_+^{out}=\frac{\kappa+i\omega}{\kappa-i\omega}\delta A_+^{in},\end{equation} one readily shows that the quadrature noise spectra of the outgoing initial modes are given by
\begin{align}
S_{X_{1,\theta}^{out}}&=\frac{|\lambda_++\lambda_-|^2}{4}S_{X_{1,\theta}^{in}}+\frac{|\lambda_+-\lambda_-|^2}{4}S_{X_{2,\theta}^{in}}+\frac{1-|\lambda_-|^2}{2},\\
S_{X_{2,\theta}^{out}}&=\frac{|\lambda_++\lambda_-|^2}{4}S_{X_{2,\theta}^{in}}+\frac{|\lambda_+-\lambda_-|^2}{4}S_{X_{1,\theta}^{in}}+\frac{1-|\lambda_-|^2}{2},
\end{align} where \begin{equation}\lambda_+=\frac{\kappa+i\omega}{\kappa-i\omega},\hspace{0.2cm}\lambda_-=\frac{\kappa+i\omega-\beta}{\kappa-i\omega+\beta}.\end{equation}

The previous relations can be straightforwardly interpreted in terms of frequency-dependent swapping. For a large enough cooperativity, $C=g^2N/2\kappa\gamma$, and not too high intensities, $\Omega'\ll g\sqrt{N}$, the intracavity fields see a cavity with an effective cavity halfwidth \begin{equation}\label{eq:kappa_CPT}\kappa_{CPT}\simeq \gamma_0+\kappa\left(\frac{\Omega'^2}{g^2N}\right),\end{equation} much narrower than the bare cavity halfwidth~\cite{lukin98,hernandez07,wu08,mucke10,kampschulte10,albert11}. One can then distinguish three regimes depending on the sideband frequency considered:
\begin{itemize}
\item[{\it (i)}] a \textit{transparency} regime for $\omega\ll\kappa_{CPT}$, where the transmission of the atom-cavity system is that of a resonant empty cavity. The fluctuations of the outgoing fields are then equal to those of the incoming fields, $\delta A_{1,2}^{out}=\delta A_{1,2}^{in}$ ($\beta\sim 0$, $\lambda_+\sim\lambda_-\sim 1$).
\item[{\it (ii)}] a \textit{swapping} regime for $\kappa_{CPT}\ll\omega\ll\kappa$, in which the dark mode sidebands see an off-resonant cavity and are therefore $\pi$-shifted, while those of the bright mode see a resonant cavity and remain unchanged. From eqs.~(\ref{eq:A-out},\ref{eq:A+out}), one thus easily obtains that $\delta A_1^{out}\sim\delta A_2^{in}$, $\delta A_2^{out}\sim\delta A_1^{in}$ ($\lambda_+\sim 1$, $\lambda_-\sim -1$).
\item[{\it (iii)}] a \textit{reflection} regime for $\omega\gg\kappa$, in which the cavity transmission is that of an off-resonant cavity and the fluctuations of the outgoing fields are those of the reflected fields, $\delta A_{1,2}^{out}=-\delta A_{1,2}^{in}$.
\end{itemize}
While the conservation of the fluctuations either within the transparency window and outside the cavity bandwidth are rather intuitive, the exchange of fluctuations in the swapping region may be less so. In this frequency window the CPT medium acts as frequency-dependent phase-plate for the field sidebands, and the dephasing is different for fluctuations of the outgoing dark and bright field modes. Indeed, while the bright mode sidebands see an empty cavity $\delta A_B^{out}\sim\delta A_B^{in}$, the dark mode sidebands see an off-resonant cavity and the intracavity field fluctuations vanish: $\delta A_D\simeq 0$, and thereby $\delta A_D^{out}\sim -\delta A_D^{in}$. This implies that the intracavity fluctuations of the initial modes are equal: $\delta A_1\sim\delta A_2$. This effect is reminiscent of the matched pulse propagation discussed in~\cite{harris93,fleischhauer94} and of the oscillatory transfer of squeezing discussed in~\cite{barberis07,hu11}, which occur with fields propagating in single-pass through a medium. However, the cavity interaction sets here a natural frequency boundary, namely the cavity bandwidth, for the atomic-induced interference effects and provides an automatic locking mechanism for the coherent exchange of fluctuations. When the atomic absorption is negligible the cavity containing the CPT medium thus plays the role of a lossless frequency-dependent phase-plate for the quantum fluctuations of the fields.
\begin{figure}[h]
\includegraphics[width=\columnwidth]{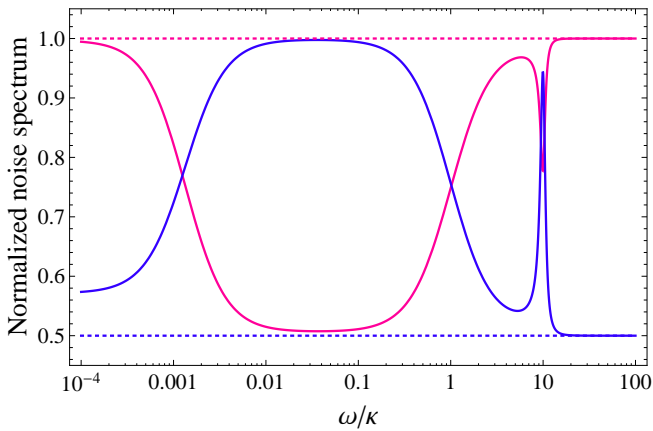}
\caption{(Color online) Noise spectra of the amplitude ($\theta=0$) quadrature of the outgoing fields ($A_1$: red, $A_2$: blue). The dashed lines indicate the amplitude-quadrature noise spectra of the incident fields, which are in a coherent and a 3dB amplitude-squeezed state, respectively ($S_{X_{1,\theta=0}^{in}}=1$, $S_{X_{2,\theta=0}^{in}}=0.5$). Parameters: $(g\sqrt{N},\gamma,\Omega,\gamma_0)=(10,0.25,0.5,0)\times\kappa$.}
\label{fig:switch_spectrum_no_gamma_0}
\end{figure}

In order to illustrate this behavior we choose the two input fields to have equal intracavity Rabi frequencies and to be in a coherent and a squeezed state, respectively. Without loss of generality, we assume field 2 to be in a broadband, minimal uncertainty, amplitude-squeezed state with a squeezing arbitrarily fixed to -3 dB: $S_{X_{2,\theta=0}^{in}}(\omega)=1/2$ and $S_{X_{2,\theta=\pi/2}^{in}}(\omega)=2$. The spectra of the outgoing fields are represented in fig.~\ref{fig:switch_spectrum_no_gamma_0}, taking as an example typical experimental parameters for cold atoms in a low-finesse cavity~\cite{lambrecht96,josse03}. For $\kappa=2\gamma$, $C=100$ and a Rabi frequency $\Omega'\sim \gamma/\sqrt{2}$ the effective cavity bandwidth is indeed much smaller than $\kappa$ $(\kappa_{CPT}/\kappa\simeq 0.001)$. As expected from the previous analysis, while the initial quadrature squeezing of field 2 is preserved both within the transparency window and outside the cavity bandwidth, it is almost perfectly transferred to field 1 in the swapping region. A symmetric behavior is observed for all quadratures. Note that the sum of the noise spectra is almost conserved at almost all analysis frequencies - at the exception of the atom-cavity normal modes ($\omega_{\pm}/\kappa\simeq \pm\sqrt{C}\simeq \pm 10$ in this case).
\begin{figure}[h]
\center{\includegraphics[width=0.9\columnwidth]{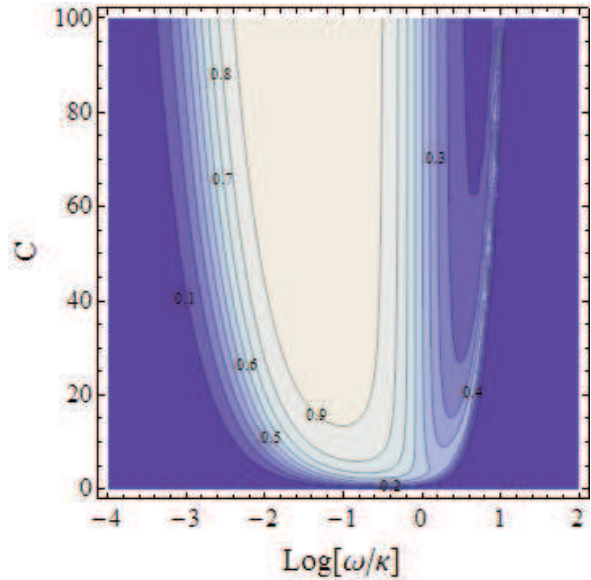}}
\caption{(Color online) Efficiency $\eta$ of the squeezing transfer from field 2 to 1, as a function of analysis frequency $\omega$ (in units of $\kappa$, log scale) and cooperativity $C$, in the same configuration as in Fig.~\ref{fig:switch_spectrum_no_gamma_0}. Parameters: $(\gamma,\Omega,\gamma_0)=(0.5,0.25,0)\times\kappa$.}
\label{fig:eta}
\end{figure}
For the same configuration of a coherent and a squeezed input fields, fig.~\ref{fig:eta} shows the efficiency of the squeezing transfer, \begin{equation} \eta\equiv\frac{1-S_{X^{out}_{1,\theta=0}}}{1-S_{X^{in}_{2,\theta=0}}}=\frac{|\lambda_+-\lambda_-|^2}{4},\end{equation} as a function of the sideband frequency $\omega$ and the cooperativity parameter $C$. In the swapping region, the efficiency rapidly increases with $C$ and can be shown to scale as \begin{align} \eta &\simeq \frac{\kappa^2}{\kappa^2+\omega^2}\frac{\omega^2}{\omega^2+\kappa_{CPT}^2}\\ &\simeq \left(\frac{1}{1+\kappa_{CPT}/\kappa}\right)^2\hspace{0.4cm}(\omega\sim \sqrt{\kappa\kappa_{CPT}}).\end{align}

We checked the effect of the ground-state decoherence using the full numerical simulations. A non-negligible $\gamma_0$  has two effects: first, it induces a coupling between the dark and bright states, thus adding excess atomic noise at low sideband frequencies (as can be seen e.g. from eq.~(\ref{eq:F-})). This excess atomic noise can lead to the reduction or disappearance of the squeezing at low frequencies. Secondly, it reduces the atomic coherence between the ground-states, thereby decreasing the quantum interference effects. We checked however that the swapping efficiency remained high as long as the transparency window is much larger than the ground-state decoherence rate. Generally, since the transparency window is ultimately limited by the ground-state decoherence rate $\gamma_0$, using long-coherence time ensembles in low-finesse cavities is thus preferable for obtaining a high efficiency as well as a large dynamical range for the swapping.

\begin{figure}[h]
\center{\includegraphics[width=0.9\columnwidth]{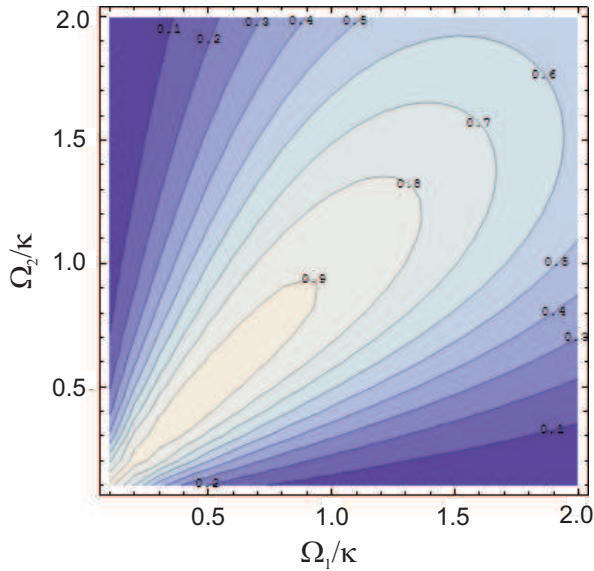}}
\caption{(Color online) Efficiency $\eta$ of the squeezing transfer from field 2 to 1, as a function of the field Rabi frequencies $\Omega_1$ and $\Omega_2$ (in $\kappa$ units). Parameters: $(g\sqrt{N},\gamma,\gamma_0,\omega)=(10,0.5,10^{-5},0.1)\times\kappa$.}
\label{fig:asym}
\end{figure}
We also examined numerically the non-symmetric situation of fields with different Rabi frequencies $\Omega_1\neq\Omega_2$. For sufficiently strong driving on both transitions the dark and bright atomic state become \begin{equation}|D\rangle=(\Omega_2|1\rangle-\Omega_1|2\rangle)/\Omega',\hspace{0.2cm} |B\rangle=(\Omega_2|1\rangle+\Omega_1|2\rangle)/\Omega'\end{equation} ($\Omega'=\sqrt{\Omega_1^2+\Omega_2^2}$), with corresponding dark and bright field combinations. Figure.~\ref{fig:asym} shows the efficiency of the squeezing transfer from field 2 to 1 as the respective Rabi frequencies of the two fields are varied. The other parameters are the same as previously and the efficiency was obtained numerically with a full calculation. Similar transparency windows are observed as in the balanced Rabi frequency case, and efficient transfer is observed as long as the CPT window is larger than the ground state decay rate and smaller than the bare cavity halfwidth. On also finds that the transfer of quantum fluctuations is most efficient for fields with balanced Rabi frequencies $(\Omega_1\sim\Omega_2)$. Qualitatively, this can be explained by the fact that the CPT-induced atomic coherence is maximal in this case and induces perfectly destructive interference for the dark mode sidebands. In an unbalanced situation the fluctuations are only partially exchanged between the initial fields. Similarly to the free-space interaction~\cite{hu11}, one can show, by following e.g. the method of~\cite{josse04}, that the fluctuations of the initial field modes can be retrieved in suitable combination of modes, which are however different from the initial ones.

To conclude, we propose to use a coherent atomic medium in an optical cavity to achieve passive quantum state swapping between two optical fields. Efficient exchange of quantum fluctuations can be achieved for reasonable effective optical depth in the bad cavity limit and when there is an appreciable narrowing of the cavity linewidth due to CPT. In addition to quantum information processing in the optical domain, the proposed mechanism could also have valuable applications for circuit QED in the microwave domain, e.g. with superconducting artificial atoms~\cite{kelly10,bianchetti10}.

\acknowledgments
The author is grateful to Michel Pinard for initiating this work and Ian D. Leroux for pointing out the phase-plate interpretation of the swapping, and acknowledges financial support from the European STREP and ITN projects \textit{PICC} and \textit{CCQED}.

\end{document}